\documentclass[letterpaper,floatfix,superscriptaddress,twocolumn]{revtex4}
\def\av#1{\left\langle#1\right\rangle}
\usepackage{latexsym}
\usepackage{amstext,amsfonts}
\usepackage{epsfig}

\newcommand{\xxsize} {14cm}

\begin{document}
\title{Catastrophic cascade of failures in interdependent networks}
   
\author{Sergey V. Buldyrev} \altaffiliation[Also at ]{Center for Polymer
  Studies and Dept. of Physics, Boston Univ., Boston, MA 02215 USA}
\affiliation{Department of Physics, Yeshiva University, 500 West 185th
  Street, New York, New York 10033, USA}

\author{Roni Parshani}
\affiliation{Minerva Center and Department of Physics, Bar-Ilan
  University, 52900 Ramat-Gan, Israel}

\author{Gerald Paul}

\affiliation{Center for Polymer Studies and Department of Physics,
  Boston University, Boston, Massachusetts 02215, USA}

\author{H. Eugene Stanley}

\affiliation{Center for Polymer Studies and Department of Physics,
  Boston University, Boston, Massachusetts 02215, USA}

\author{Shlomo Havlin}

\affiliation{Minerva Center and Department of Physics, Bar-Ilan
  University, 52900 Ramat-Gan, Israel}

\date{7 July 2009 revision --- mutual.tex}

\begin{abstract}
  Many systems, ranging from engineering to medical to
  societal, can only be properly characterized by multiple
  interdependent networks whose normal functioning depends on one
  another. Failure of a fraction of nodes in one network may lead to a failure in another network. 
  This in turn may cause further malfunction of additional nodes in the first network and so on.
  Such a cascade of failures, triggered by a failure of a small faction of nodes in only
  one network, may lead to the complete fragmentation of all networks. 
  We introduce a model and an analytical framework for studying interdependent networks.
  We obtain interesting and surprising results that should significantly effect the design of robust real-world networks.
  For two interdependent Erdos-Renyi (ER) networks, we find that the critical average degree below which both  
  networks collapse is $\langle k_c \rangle=2.445$, compared to $\langle k_c \rangle=1$ for a single ER network. 
  Furthermore, while for a single network a broader degree distribution of the network nodes results in higher robustness to random failure, 
  for interdependent networks, the broader the distribution is, the more vulnerable the networks become to random failure.
  
\end{abstract}

\maketitle

\section{Introduction}

After a decade of intense study on networks, almost all work done has 
concentrated on the limited case of a single network which does not interact with other networks \cite{ws,barasci,bararev,PastorXX,mendes,cohena,vespig,Newman-handbook,mnewman,Callaway,song,boguna}. 
Such situations rarely, if ever, occur in nature. Just as in the case of idealized gas, when interactions are present as in nature, 
new physical laws appear. 

Analogously, due to technological progress, modern systems are becoming more and more coupled together.
While in the past many networks would provide their functionality independently, modern systems depend on
one another to provide proper functionality.
For example, a power network in which the nodes are power stations and a communication network 
in which the nodes are computers, are interdependent, since nodes from the communication network rely for power supply on the 
power stations, while the power stations depend for their control on the proper functioning of the communication network.
The critical importance of functional dependence of networks on each
other has recently been recognized \cite{chiaradonna, laprie}.

In interdependent networks, when nodes in one network fail, they cause dependent nodes in another network to fail.
This may happen recursively and can lead to a cascade of failures. So a failure of a small faction of nodes in only one network, may lead to the complete fragmentation of all networks. 
Here, we provide a framework for understanding the robustness of interacting networks subject to such cascading failures and 
provide the basic network analytic approach that can underlie future work in this area. 
We present a general model for interdependent networks that we solve analytically using tools from percolation theory and the apparatus of
generating functions. We present exact analytical solutions for the critical fraction of nodes that upon removal will lead to a 
complete fragmentation of all networks.

Surprisingly, analyzing complex systems as a set of interdependent networks may destabilize the most basic assumptions 
that network theory has relied on for single networks. While for a single network a broader degree distribution of the network nodes results in the network being more robust to random failure, for interdependent networks, the broader the distribution is, the more vulnerable the networks become to random failure. The implications are dramatic -- the current methods applied to the design of robust networks need to be modified to include the findings about interdependent networks.     

\section {The model}

Consider two networks A and B and assume that the functioning of a
node in network A depends on the ability of one or more nodes in
network B to supply a critical resource to the node in network A.
Similarly, a node in network B depends on a set of nodes in network A.
The networks can be connected in different ways; in the most general configuration
one could specify the distributions of connections between the nodes from both networks.
 
The networks can have the same, or different, topologies. 
The model can easily be extended to an arbitrary number of interacting
networks each with its own specific topology and dependence on the
other networks. For example, an interesting dependence for three interacting networks could be a 
circular dependency in which the nodes in network B depend on network A for a resource, the
nodes of network C depend on the nodes of network B for a resource and
the nodes of network A depend on network C for resources.

Our key insight from percolation theory is that for each of the
networks to remain functional after nodes have failed, the network
must include a spanning cluster of functional nodes.  Nodes that are not
part of the spanning cluster will become nonfunctional and will cause the nodes from other networks that are connected to them 
to also become nonfunctional.

For simplicity and without loss of generality, we will assume a system
of two networks, A and B, both with N nodes.  Within each network, we
assume that nodes are randomly and independently connected with degree
distributions $P_A(k)$ and $P_B(k)$.  We also assume, for simplicity, that each node in
network A is connected to, and dependent on, one node in network B and
vice-versa.  Next we will remove a fraction of nodes $1-p$ from
network A and all the edges connected to these nodes, so that only a
fraction $p$ remain functional.  Simultaneously, the corresponding
nodes (and their edges) in network B are removed since they are
dependent on the nodes in network A.

As edges are removed, the networks break up into connected
components (clusters).  The clusters in network A and the clusters in
network B will be different because the networks are each connected
differently.  We define a mutually-connected cluster as the set of
nodes in network A which belong to a cluster in network A and also
have their corresponding nodes in network B belong to a single cluster
in network B (or vice-versa). We assume that clusters of nodes that are
disconnected from the network core (giant component/spanning cluster)
become non-functional and are removed. Thus, the mutually-connected
giant component will be of special interest since it is the only functional part of the system.

The questions that we will ask: What is the critical $p=p_c$
below which all the mutual clusters constitute only an infinitesimal
fraction of the network, i.e., no mutual giant component exist? What
is the probability $p_\infty(p)$ for a node to belong to the mutual
giant component as function of $p$?  To solve this model we will
introduce a recursive process which we will identify with a physically
meaningful cascade of failures.

To solve this model we will first define the $a_1$-clusters of network A
after only a fraction of nodes $p$ remain. Next we will treat each of
these $a_1$-clusters as separate subsets of a network B, i.e. all the
B-edges connecting different $a_1$-clusters will be removed. We will
define this state of the networks as the first stage in the cascade of
failures. Accordingly, each of the $a_1$-clusters may split into several
$b_2$-clusters. Some of the $a_1$-clusters will not split and will
coincide with $b_2$ clusters. Obviously such clusters are mutually
connected. Finally we remove from the network all $A$-edges connecting
different $b_2$ clusters. We will define this state of the networks as
the second stage in the cascade of failures. Analogously, in the third
stage we will determine all the $a_3$-clusters and in the fourth stage
we will determine all the $b_4$-clusters, and will continue this process
until no further splitting and edge-removal will occur.

Note that in this process the majority of new mutual clusters identified
after each stage of failures will be isolated nodes, few of them will be
of size 2 and very rarely we will have larger mutual clusters. Indeed if
we have two nodes that are connected by an A-edge, the probability that
they will be connected by a $B$ edge is $1-\sum_k P_B(k)(1-1/N)^k\approx
1-\sum_k P_B(k)(1-k/N)= \sum_k P_B(k)k/N=<k>_B/N \to 0$ for $N\to
\infty$. The probability that three nodes connected by A-edges are also
connected by $B$ edges scale as $1/N^2$ and so on. 

\section{Analytical solution}

Now we will solve the problem analytically using the apparatus of
generating functions. As in Refs.~\cite{Newman,Shao} we will introduce
generating functions of the degree distributions
\begin{equation}
G_{A0}(x)=\sum_k P_A(k) x^k 
\end{equation} 
and
\begin{equation}
G_{B0}(x)=\sum_k P_B(k) x^k. 
\end{equation} 
Analogously we will introduce generating functions of the underlining branching processes:
\begin{equation}
G_{A1}(x)=G'_{A0}(x)/ G'_{A0}(1)
\end{equation} 
and
\begin{equation}
G_{B1}(x)=G'_{B0}(x)/ G'_{B0}(1).
\end{equation}

Random removal of fraction $1-p$ of nodes will change the degree
distribution of the remaining nodes \cite{Newman}, so that the new
generating functions become
\begin{equation}
G_{A0}(x,p)=G_{A0}(1-p(1-x)),
\end{equation}
\begin{equation}
G_{B0}(x,p)=G_{B0}(1-p(1-x)),
\end{equation}
\begin{equation}
G_{A1}(x,p)=G_{A1}(1-p(1-x)),
\end{equation}
and
\begin{equation}
G_{B1}(x,p)=G_{B1}(1-p(1-x)).
\end{equation}
Let us denote the subset of nodes after the random removal of $1-p$
nodes as $A_0=B_0\subset A=B$. If the number of nodes in the entire network
is $N$, the number of nodes in $A_0=B_0$ is $N_0=pN$.  The fraction of
nodes that belong to the giant component of the network $A_0$ is
\cite{Shao}
\begin{equation}
p_A(p)=1-G_{A0}(f_A,p),
\label{e:p_A} 
\end{equation}
where $f_A$ satisfies a transcendental equation
\begin{equation}
f_A(p)=G_{A1}(f_A,p).
\label{e:f_A}
\end{equation}
Equation (\ref{e:f_A}) can be simplified by substitution $z_A=1-p(1-f_A)$
\begin{equation}
1-1/p+z_A/p=G_{A1}(z_A).
\label{e:z_A} 
\end{equation}
Then Eq. (\ref{e:p_A}) becomes 
\begin{equation}
p_A(p)=1-G_{A0}(z_A),
\label{e:pz_A} 
\end{equation}

Analogous equations characterize the giant component of network $B_0$.
After the initial attack which removes $(1-p)$-fraction of nodes from
both networks, the first-stage failure is caused by the fragmentation of
the subset $A_0$. The giant component $A_1$ of $A_0$ will constitute
$P_A(p)$-fraction of $A_0$. Thus the number of nodes in $A_1$ is
$N_1=N_0P_A(p)=pP_A(p)N=p_1N$.

After the first-stage failure the fraction of functioning nodes is
$p_1=pP_A(p)$ (subset $A_1$). Because the nodes of the networks B and A
coincide, the same fraction of nodes remains functioning in network
B. Because the topology of network B is independent the topology of
network A, these functioning nodes are totally random with respect to
connections in network B.  Thus we can again apply the apparatus of
generating functions and find the fraction $P_B(p_1)$ of the giant
component $B_2$ of network B with respect to the subset $A_1$. The
number of nodes in the giant component $B_2\subset A_1$ is $N_2\equiv
p_2N=P_B(p_1)N_1= P_B(p_1)p_1N=pP_A(p)P_B(p_1)N$. Thus, the fraction of
functioning nodes after the second stage failure is
$p_2=pP_A(p)P_B(p_1)$ (subset $B_2$).

Now we will analyze what happens during the third-stage failure which is
caused by further fragmentation of the giant component $A_1$ by removal
$N_1-N_2=(1-P_B(p_1)N_1$, nodes which do not belong to $B_2$. The
removal of these nodes form $A_1$ is equivalent to the
removal of the same fraction of nodes from $A_0$ (because all the nodes
that were removed at the stage of the initial attack do not belong to
$B_2$, $A_1$, and $A_0$. The total number of nodes that must be removed
from network A is $(1-P_B(p_1))N_0$ nodes from $A_0$ plus the number of
the initially attacked nodes $(1-p)N$. Thus, the total number of nodes
that must be removed from network A is $(1-pP_B(p_1))N$.  Hence the
third-stage failure is equivalent to a random attack in which $p$ is
replaced by $p'_2=pP_B(p_1)$. Accordingly the number of nodes in the
giant component $A_3\subset B_2$ is $N_3\equiv p_3N=p'_2P_A(p'_2)$.

Following this approach we can construct the sequence of giant components in
the cascade of failures: $A_{2m+1}\subset B_{2m}\subset
A_{2m-1}\ldots\subset A_3\subset B_2 \subset A_1 \subset B_0=A_0 \subset
A=B$.  The number of modes in each giant component of this sequence is
$N>pN\equiv Np_0>N p_1 >\ldots Np_{2m+1}\ldots$, where the numbers $p_n$
can be obtained by recursive relations: $p_0\equiv p'_0\equiv p$,
$p_1\equiv p'_1\equiv p'_0 P_A(p'_0)$, $p'_2=p P_B(p'_1)$,
$p_2=p'_1P_B(p'_1)$, $p'_3=p P_A(p'_2)$, $p_3=p'_2P_A(p'_2)\ldots
p'_{2m}=p P_B(p'_{2m-1})$, $p_{2m}=p'_{2m-1}P_B(p'_{2m-1})$,
$p'_{2m+1}=p P_A(p'_{2m})$, $p_{2m+1}=p'_{2m}P_A(p'_{2m})$.

Now we will determine the size of the mutual giant component.  The
fraction of nodes in the mutual giant component, $P_\infty$ is the limit of the sequence $p_n$  for $n\to\infty$.  This limit must satisfy the equations
$p_{2m+1}=p_{2m}=p_{2m-1}$
since the cluster is not further fragmented.
Using relations between $p_{n}$ and $p'_{n-1}$, and denoting
$p'_{2m-1}=x$ and $p'_{2m}=y$ we arrive to a system of two symmetric
equations with two unknowns:
\begin{equation}
\left\{    
\begin{array}{lr}
x=pp_A(y)\\
y=pp_B(x).
\end{array}
\right.\
\label{system}
\end{equation}
This system of equations has one trivial solution $x=0$, $y=0$ for any
$p$, corresponding to the zero size of the giant mutual component. If
$p$ is large enough there exists a different solution which gives the
nonzero size of the mutual giant component.  We can easily exclude $y$
from these equations and obtain a single equation 
\begin{equation}
x=pp_A(pp_B(x))
\label{e:xpp}
\end{equation}
This
equation can be solved graphically~(Fig.\ref{f:1}) 
as the intersection of a straight
line $y=x$ and a curve $y=pp_A(pp_B(x))$ which both intersect at the
origin. When $p$ is small enough the curve increases very slowly and
does not intersect with the straight line. The critical case when the
nontrivial solution emerges, corresponds to the case when the line
touches the curve at a single point $x$ and in this point we have a
condition $1=p^2p'_A(pp_B(x))p'_A(x)$, which together with equation
$x=pp_A(pp_B(x))$ gives the solution for the critical $p$ and the critical size
of the mutual giant component.

\section{ER networks}

In case of ER networks, whose degrees are Poisson-distributed \cite{er1,bollo}, the
problem can be solved explicitly. Suppose that the average degree of the
network A is $a$ and the average degree of the network B is $b$.  Then, 
$G_{A1}(x)=G_{A0}=\exp(a(x-1))$ and $G_{B1}=G_{B0}=\exp(b(x-1)$.
Accordingly system (\ref{system}) becomes
\begin{equation}
\left\{    
\begin{array}{lr}
x=p[1-f_A]\\
y=p[1-f_B],
\end{array}
\right.\
\label{systemf0}
\end{equation}
where 
\begin{equation}
\left\{    
\begin{array}{lr}
f_A=\exp[ay(f_A-1)]\\
f_B=\exp[bx(f_B-1)].
\end{array}
\right.\
\label{systemf1}
\end{equation}
Excluding $x$ and $y$, we get a system with respect to $f_A$ and $f_B$:
\begin{equation}
\left\{    
\begin{array}{lr}
f_A=e^{-ap(f_A-1)(f_B-1)}\\
f_B=e^{-bp(f_A-1)(f_B-1)}.
\end{array}
\right.\
\label{systemf2}
\end{equation}
Introducing a new variable $r=f_A^{1/a}=f_B^{1/b}$, we reduce system
(\ref{systemf2}) to a single equation
\begin{equation}
r=e^{-p(r^a-1)(r^b-1)},
\label{systemr}
\end{equation}
which can be solved graphically for any $p$. The critical case
corresponds to the tangential condition
\begin{equation}
1=\frac{d}{dr}e^{-p(r^a-1)(r^b-1)}=p[ar^a+br^b-(a+b)r^{a+b}],
\label{tg}
\end{equation}
from where the critical value of $r=r_c$ satisfies transcendental equation
\begin{equation}
r=e^{-\frac{(1-r^a)(1-r^b)}{ar^a+br^b-(a+b)r^{a+b}}},
\label{rc}
\end{equation}
and the critical value of $p=p_c$ can be found from Eq.~(\ref{tg}).
\begin{equation}
p_c=\frac{1}{ar_c^a+br_c^b-(a+b)r_c^{a+b}}.
\end{equation}
The values of $p_c$ and $P_\infty$ for different $a$ and $b>a$ are 
presented in Fig.~\ref{f:2} as function of $a/b$.
In case $a=b$, $f_A=f_B=f$, and $f_c$ satisfy equation
\begin{equation}
f_c=e^{-\frac{(1-f_c)^2}{2f_c^2+2f_c}},
\label{fc}
\end{equation}
which gives a solution $f_c=0.28467$, $p_c=2.4554/a$, and the critical
fraction of nodes in the mutual giant component
$P_\infty=p_c(1-f_c)^2=1.2564/a$.  Numerical simulations of the ER
networks are in excellent agreement with the theory (Fig.~\ref{f:3}).

\section{Scale-free networks}

For regular percolation in a scale-free network with a power
law degree distribution $P_A(k)\sim k^{-\lambda_A}$, it is known that $p_c\to 0$, as
$N\to \infty$ for $\lambda_A\leq 3$. Surprisingly, for mutual percolation this is not
the case and $p_c$ remains finite for $\lambda_A>2$. To see
this, we can find analytical approximation for $P_A(p)$. First, we begin by solving Eq.(\ref{e:z_A}). According to Tauberian
theorems, for $\lambda_A\leq 3$, $G_{A1}(x)$ has a singularity at
$x=1$ of the sort $1-\kappa_A(1-x)^{\lambda_A-2}$.  Therefore it
has a diverging derivative which has a physical meaning of the
branching factor $\tilde {k}_A$. To solve Eq. (\ref{e:z_A}), we must
find the intersection of the straight line $y=1-1/p+z/p$ and
the curve $y=G_{A1}(z)$. The straight line passes through the point $y=1,z=1$ with
the derivative $1/p$. Thus there is always a trivial solution $z=1$,
which corresponds to the absence of percolation. If,
$G'_{A1}(1)=\tilde{k}_A$ is finite, we do not have another solution for
$p<1/\tilde{k}_A$ (a classical result for regular percolation), but for
$\lambda_A\leq 3$, we always have a non trivial solution
$z=1-(p\kappa_{A})^{1/(3-\lambda_A)}$. Since $G'_{A0}(1)=\langle k
\rangle$, which is finite for $\lambda_A>2$, Eq. (\ref{e:pz_A}) 
yields
$P_A(p)=(p\kappa_A)^{1/(3-\lambda_A)}\langle k_A\rangle$.
Finally Eq.(\ref{e:xpp}) becomes
\begin{equation}
x=p\langle k_A\rangle\left[p\kappa_A \langle k_B\rangle (\kappa_Bx)^{1/(3-\lambda_B)}\right]^{1/(3-\lambda_A)}.
\end{equation}
The right-hand side of this equation behaves as $x^\mu$, where
$\mu=1/[(3-\lambda_A)(3-\lambda_B)]>1$. Thus the  r.h.s curve always 
goes below $y=x$ for $x\to 0$, so for
sufficiently small $p$ we do not have a non-trivial solution, which means
the absence of the mutual giant component. Thus we have
a percolation transition at some $p=p_c>0$.

\section{Robustness of interdependent networks}

For interdependent networks we find the surprising behavior that networks with 
a broad degree distribution (of the network nodes) are more vulnerable to random attack compared to networks with a narrow distribution.
To understand this result we note the following:  
1) All interdependent networks are randomly connected, high degree nodes from one network might be connected to low degree 
nodes from the other networks.
2) At each step when nodes (and their links) are disconnected from one network their corresponding nodes (and their edges) from 
the other network are also removed.

Therefore, the hubs that play such a dominant role in the robustness of single networks become vulnerable when a cascade of failures occurs in 
interdependent networks. Moreover, for a network with a fixed average degree, a broader distribution means more nodes with low degree to balance the high degree nodes. Since the low degree nodes are more easily disconnected the advantage of a broad distribution on single networks becomes a disadvantage for several interdependent networks.       

In Fig.~\ref{avDegDistribution} we compare simulation results for several SF networks with different $\lambda$ values, an ER network and a Random Regular (RR) network, all with an average degree $\av{k}=4$. The simulation results are in full agreement with our analytical results and it can clearly be seen that for a broader distribution $p_c$ is indeed higher.

\section{Finite size effects}

Our considerations are rigorous for $N\to\infty$. For a finite network,
the relative fluctuations of all fractions decrease as $1\sqrt{N}$ so,
for the finite network, there is a range of values of $p$ for which the
mutual giant component exists with probability $P_\infty(p)$ (Fig.~\ref{f:4}). 
Its derivative diverges as $N\to\infty$ as $dP_\infty/dp\vert_{p=p_c}\sim
\sqrt{N}$, and for $N\to\infty$, $P_\infty(p)$ becomes a step function
$P_\infty(p)=0$ for $p<p_c$ and $P_\infty(p)=1$ for $p>p_c$. 
The square-root scaling with $N$ of the width of the interval $p$ for which we can
have a complete fragmentation for some realizations of networks and
a giant component for the other realizations of the networks can be justified
by the following arguments. The actual fraction of the remaining
nodes $p_a$ in a finite network of size $N$ will be normally distributed
around given $p$ with the standard deviation inverse proportional to $N$. 
Thus $P_\infty(p)$ is equal to  
the probability that $p_a>p_c$, which is equal to the integral of the normal 
probability density with zero mean and the same standard deviation from $p_c-p$ to infinity. Therefore the derivative  $dP_\infty/dp$ has a Gaussian
shape with standard deviation proportional to $1/\sqrt{N}$. 

The average number of stages $\langle n \rangle$ in a cascade of failures for 
$p>p_c$ diverges proportionally 
to $\ln N/\sqrt{p-p_c}$. This follows from the properties of the 
iterative process Eq. (\ref{e:xpp}). This can be seen from the fact
that near $p=p_c$, Eq. (\ref{e:xpp}) has two roots produced by the 
intersection of the curvy line which can be approximated by a parabola
$y=a(p)x^2+b(p)x +c(p)$ 
and a straight line $y=x$ (Fig.~\ref{f:1}). This is equivalent to solving a quadratic equation
$a(p)x^2+(b(p)-1)x+c(p)=0$. The value $p=p_c$ is given by the discriminant of this 
equation equal to zero: $d(p_c)\equiv (b(p_c)-1)^2-4a(p_c)b(p_c)=0$. In the general case, 
all three parameters, $a(p)$, $b(p)$, and $c(p)$, 
have non-zero derivatives at $p=p_c$. Therefore, in the general case $d(p)$ has
also a non-zero derivative at $p=p_c$, and hence the difference between the
roots scales as $\sqrt{p-p_c}$. Thus, 
the derivative of the curve at the largest root, which corresponds to the limit of the iterative process scales as $f'=1-\alpha\sqrt{p-p_c}$, where $\alpha$
is some positive constant. For Eq.(\ref{system}) the iterations converge to the root as $f'_n=\exp(-\alpha\sqrt{p-p_c}n)$. In a real network, they will stop when
the difference between two successive iterations will be smaller than one
node, which yields a condition $\exp(-\alpha\sqrt{p-p_c}n) \sim 1/N$. Hence
indeed $\langle n \rangle  \sim \ln N/\sqrt{p-p_c}$. 

For $p<p_c$ the
solution does not exist and the curve misses the line with the distance
proportional to the negative descriminant. As the curve comes close to
the line the steps are proportional to $(x-x_c)^2+d$, where $d\sim p_c-p$ 
is the minimal distance between the curve and the line. The number of such 
steps per $dx$ is $dx/((x-x_c)^2+d)$. The total number of steps are thus the integral of this quantity between $x=p$ and $x=0$, which in the limit $d\to 0$ gives
$\langle n \rangle=\pi/\sqrt{d}\sim 1/\sqrt{p_c-p}$.  

Exactly at the critical point $p=p_c$ the straight line touches the
curve at a single point and the sequence of iterations converges as
$x_{n+1}-x_c=x_{n}-x_c - a (x_{n}-x_c)^2$. These iterations converge
to $x_c$ as $1/n$ which can be seen by plugging into this equation
$x_n-x_c = C/n^\beta +o(n^{-\beta})$ where $C$ and $\beta$ are some
unknown constants. Expanding $(n+1)^{-\beta}$ in Tailor series and
equating coefficients for equal powers, one can see that
$\beta=1$. However, in real network, due to Gaussian spread in $p_a$,
we are never at criticality, and the typical $p_a-p_c\sim
1\sqrt{N}$. Therefore the distributions of the number of stages in the
cascade has an exponential tail $\exp[-\alpha n\sqrt{p_a-p_c}]$, in
which $(p_a-p_c)$ must be replaced by its typical value $1/\sqrt{N}$.
Therefore, the distribution of $P(n)$ must have an exponential tail
$P(n)\sim\exp[-\alpha' n/N^{1/4}]$, where $\alpha'$ is some positive
constant.  Thus at criticality, we expect that $\langle n \rangle \sim N^{1/4}$ as supported by our simulations
(Fig.~\ref{f:5}).


\newpage

\begin{widetext}

\begin{figure}[tbh]
\centerline{
\epsfbox[80 400 700 540]{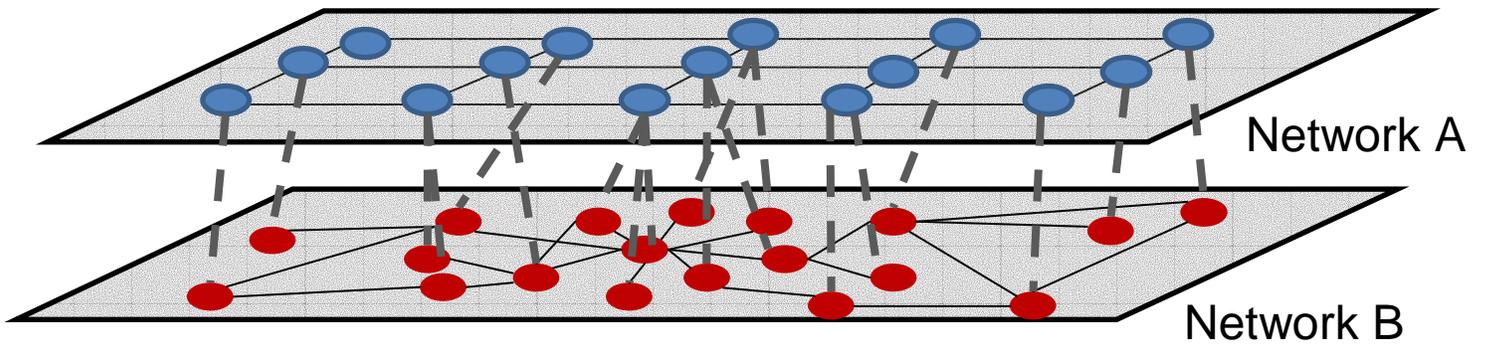}
}
\caption{(Color online) Demonstration of two interdependent networks.  Nodes in network B
  (communications network) are dependent on nodes in network A (power
  grid) for power; nodes in network A are dependent on network b for
  control information. General case is represented in which there is
  not a one-one correspondence between nodes in networks. }
\label{ex1}
\end{figure}

\begin{figure}
\includegraphics[width=\xxsize,angle=0]{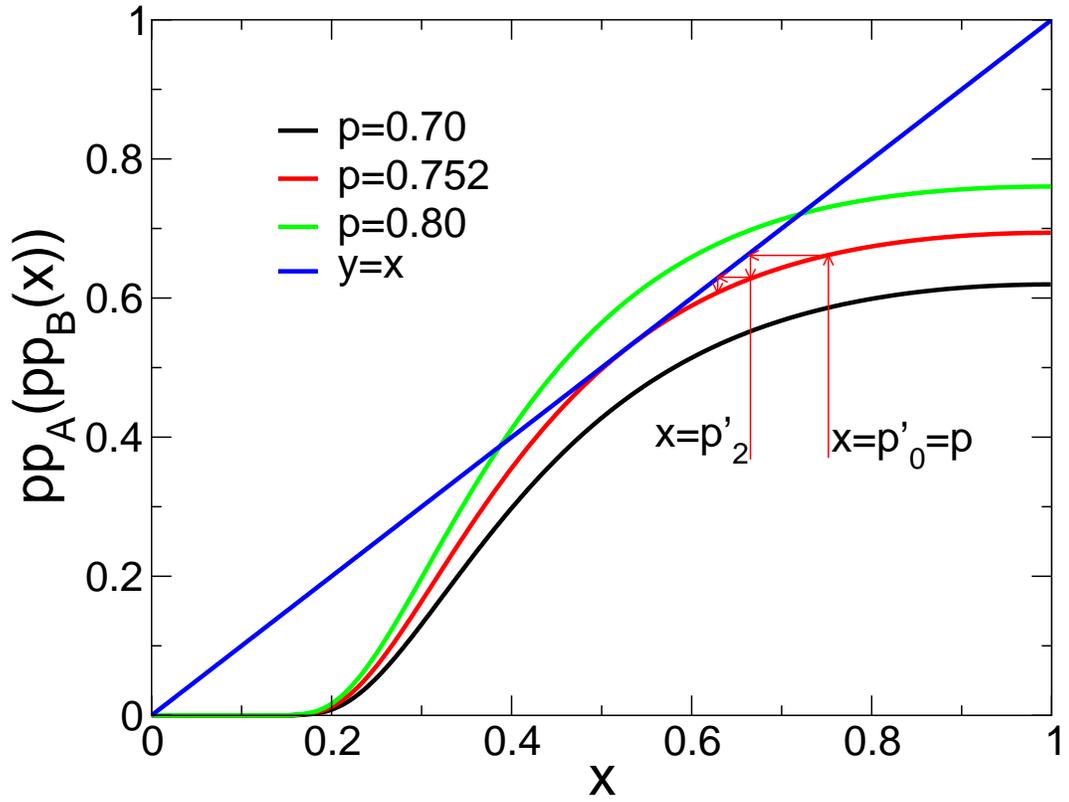}
\caption{\label{f:1} (Color online) Iterative process described 
by Eq.(\ref{e:xpp}) for the case of the scale-free distribution 
$P_A(k)=P_B(k)=(2/k)^2-(2/(k+1))^2$ for $k=2,3,...$. For $k\to \infty$, this
distribution scales as $k^{-\lambda}$, where $\lambda=\lambda_A=\lambda_B=3$.
Three curves corresponding to $p=0.70<p_c$ (black), $p=0.752\approx p_c$ (red)
and $p=0.80> p_c$ (green). One can see that for $p\geq p_c$, the iterations
(red arrows) 
starting from $p'_0=p$, converge to the largest of the two roots of 
Eq.(\ref{e:xpp}).  For $p<p_c$, the iterations converge to 0.   }
\end{figure}

\begin{figure}
\includegraphics[width=\xxsize,angle=0]{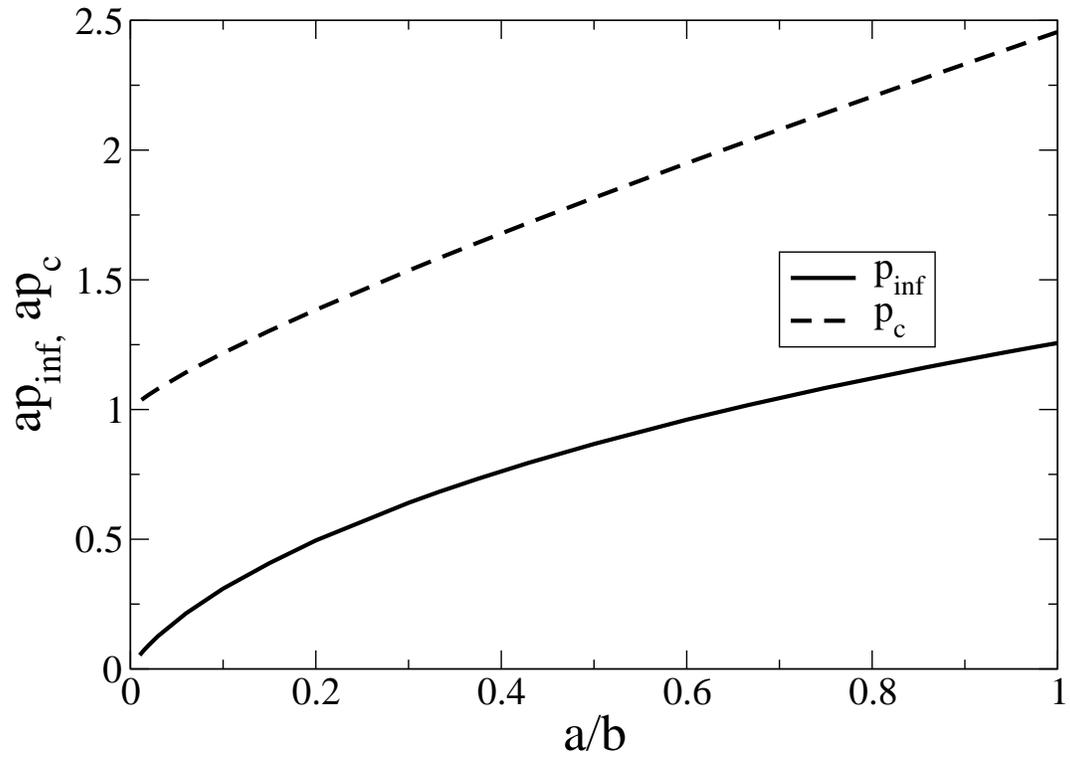}
\caption{\label{f:2} ER networks: critical fraction $p_c$ and the fraction
of nodes in the mutual giant component at criticality $P_\infty$ as function
of the ratio $a/b$, where $a$ and $b$ are the average degrees 
of networks A and B respectively.}
\end{figure}

\begin{figure} [b]
\includegraphics[width=\xxsize,angle=0]{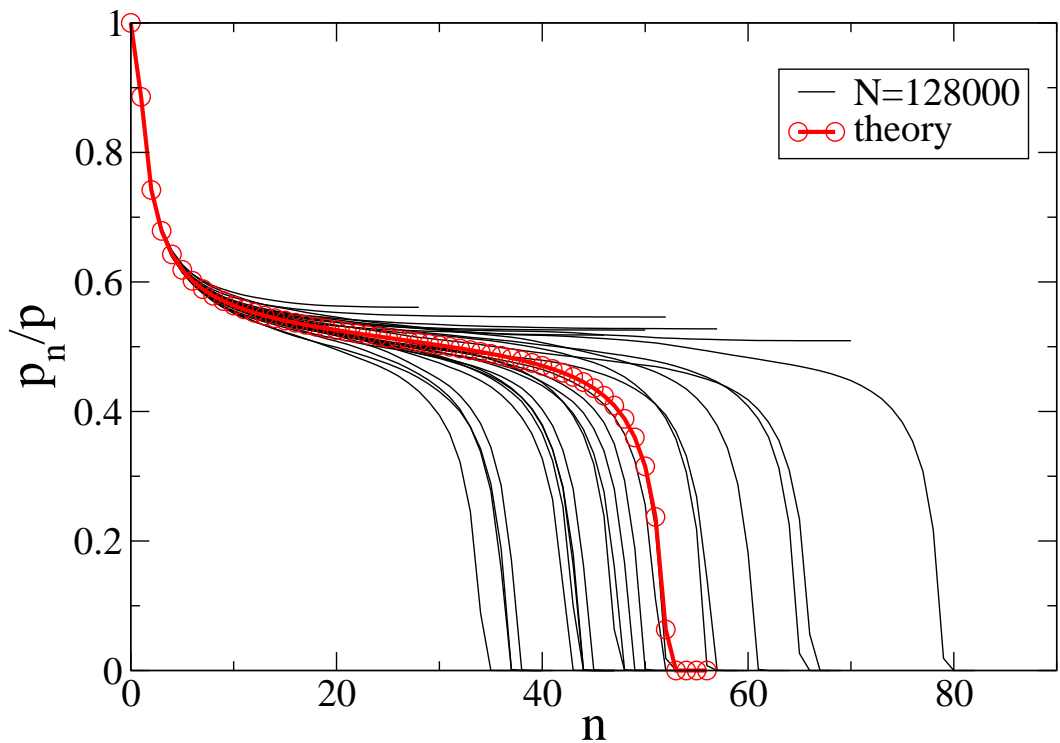}
\caption{\label{f:3} (Color online) Comparison of the fraction of the 
giant components after $n$ stages of the cascade failures for several random
realizations of ER networks
with $a=b$, $N=128000$ and $a p=2.45<a p_c=2.4554$ 
and theoretical prediction of Eq.(\ref{system}). One can see that 
for the initial stages the agreement is perfect, however at larger $n$
the deviations due to random fluctuations of the order of $1/\sqrt{N}$ 
in the actual fraction of the remaining nodes $p_a$ start to increase. 
The theoretical prediction after a region of the plateau around the 
the critical value, drops to zero, corresponding to the complete
fragmentation of the network. The random realizations separate into two
classes: one that converge to a mutual giant component and the other 
that results in a complete fragmentation.}
\end{figure}

\newpage
\begin{figure}
\includegraphics[width=\xxsize,angle=0]{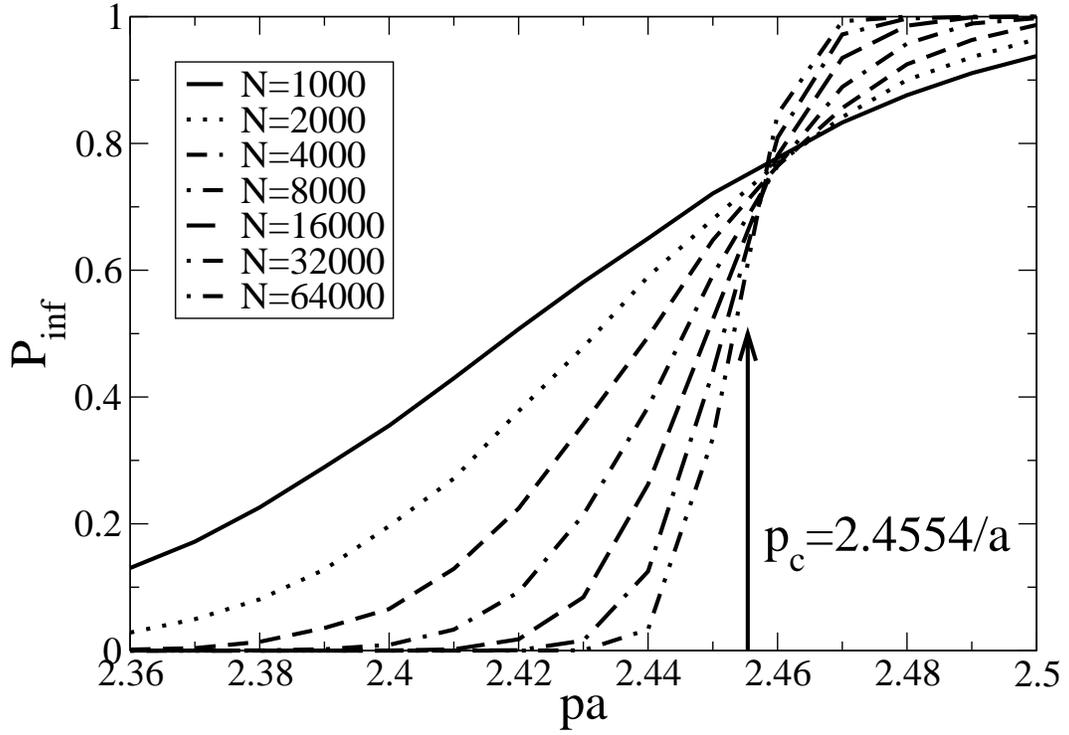}
\caption{\label{f:4} Numerical simulations of ER networks with $a=b$
and finite number of nodes, $N$. The probability of existence of
the mutual giant component $P_{\infty}$, is shown as function of $p$ for
different $N$. One can see that as $N\to \infty$ the curves converge
to a step function. The theoretical prediction of $p_c$ is shown by the arrow.}
\end{figure}

\begin{figure}
\includegraphics[width=\xxsize,angle=0]{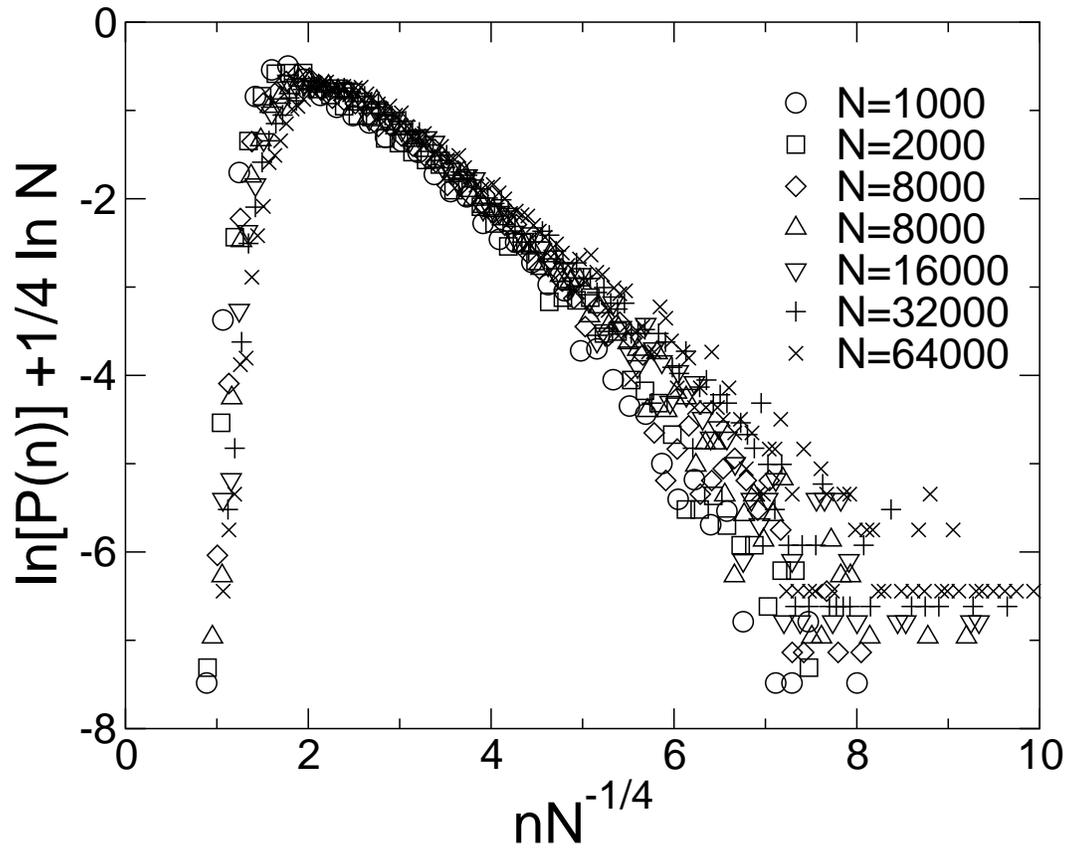}
\caption{\label{f:5} Scaled distribution of the number of stages
in the cascade failures for ER graphs with $a=b$ at criticality ($pa=2.4554$)
for different values of $N$. }
\end{figure}

\begin{figure}
\includegraphics[width=\xxsize,angle=0]{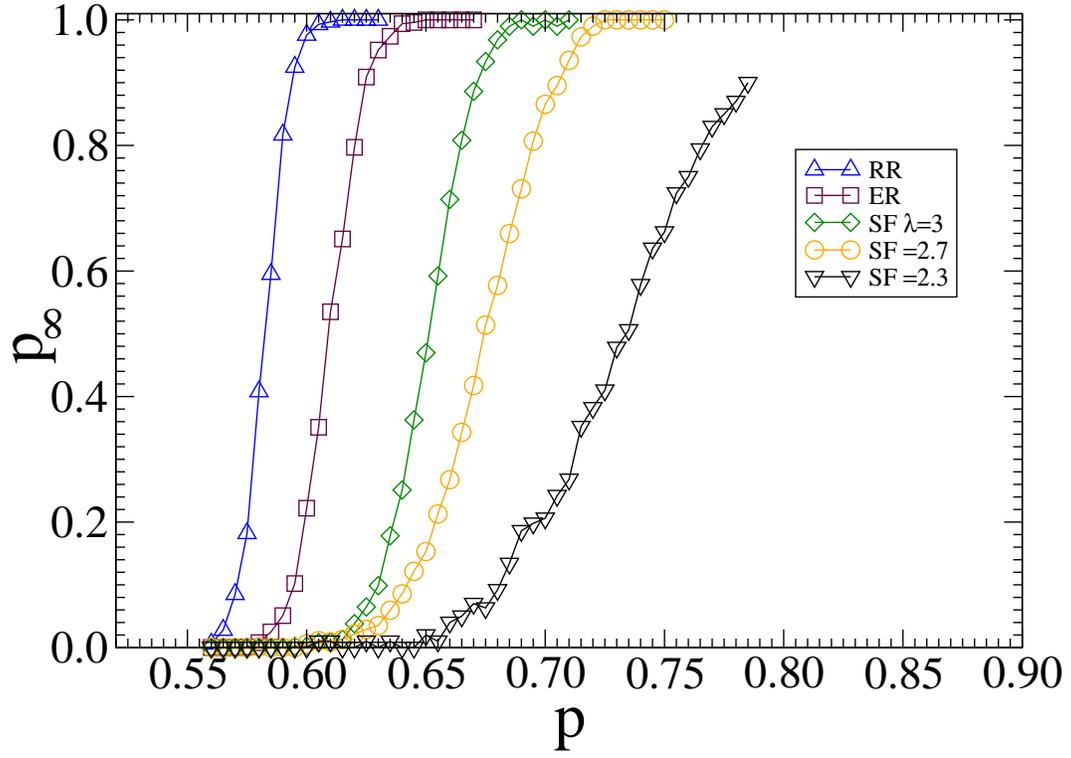}
\caption{\label{avDegDistribution} (Color online) Simulation results for $P_{\infty}$ as a function of $p$  for 
for SF networks with $\lambda = 3,2.7,2.3$, an ER network and a Random Regular (RR) network, all with an average degree $\av{k}=4$. The simulation results are with full agreement with our analytical results and it can clearly be seen that for a broader distribution $p_c$ is higher.}
\end{figure}

\end{widetext}

\end{document}